# Controlling False Positives in Association Rule Mining

Guimei Liu
School of Computing
National University of Singapore
liugm@comp.nus.edu.sg

Haojun Zhang
School of Computing
National University of Singapore
zhanghao@comp.nus.edu.sg

Limsoon Wong
School of Computing
National University of Singapore
wongls@comp.nus.edu.sg

## ABSTRACT

Association rule mining is an important problem in the data mining area. It enumerates and tests a large number of rules on a dataset and outputs rules that satisfy user-specified constraints. Due to the large number of rules being tested, rules that do not represent real systematic effect in the data can satisfy the given constraints purely by random chance. Hence association rule mining often suffers from a high risk of false positive errors. There is a lack of comprehensive study on controlling false positives in association rule mining. In this paper, we adopt three multiple testing correction approaches—the direct adjustment approach, the permutation-based approach and the holdout approach—to control false positives in association rule mining, and conduct extensive experiments to study their performance. Our results show that (1) Numerous spurious rules are generated if no correction is made. (2) The three approaches can control false positives effectively. Among the three approaches, the permutation-based approach has the highest power of detecting real association rules, but it is very computationally expensive. We employ several techniques to reduce its cost effectively.

## Categories and Subject Descriptors

H.2.8 [**DATABASE MANAGEMENT**]: Database Applications—*Data Mining*

## 1. INTRODUCTION

Association rule mining was first introduced by Agrawal et al.[2] in the context of transactional databases. It aims to find rules of the form: $X \Rightarrow Y$, where $X$ and $Y$ are two sets of items. The meaning of the rule is that if the left-hand side $X$ occurs, then the right-hand side $Y$ is also very likely to occur. The interestingness of the rules is often measured using support and confidence. The support of a rule is defined as the number of records in the dataset that contain both $X$ and $Y$. The confidence of a rule is defined as the proportion of records containing $Y$ among those records containing $X$. Association rule mining outputs rules with support no less than $min\_sup$ and confidence no less than $min\_conf$, where $min\_sup$ is called the minimum support threshold and $min\_conf$ is called the minimum confidence threshold. The two thresholds are specified by users.

An association rule implies the association between its left-hand side and its right-hand side. A question that arises naturally is how likely the association between the two sides is real, that is, how likely the occurrence of the rule is due to a systematic effect instead of pure random chance. Rules that occur by chance alone are not statistically significant. In statistics, p-value is used to measure the statistical significance of a result. In the case of association rules, the p-value of a rule $R$ is defined as the probability of observing $R$ or a rule more extreme than $R$ given the two sides of $R$ are independent. If a rule $R$ has low p-value, then $R$ has a low chance to occur if its two sides are independent. Given that $R$ is observed in the data, then its two sides are unlikely to be independent, that is, the association between them is likely to be real. A high p-value means that $R$ has a high chance to occur even if there is no association between its two sides. A rule with high p-value cannot tell us whether its two sides are dependent. Such rules should be discarded. Conventionally, a p-value of 0.05 is recognized as low enough to regard a result as statistically significant [6].

A p-value threshold of 0.05 means that there is a 0.05 probability that a rule is not real but we are wrongly regarding it as real. If we test 1000 random rules at the significance level of 0.05, then around 50 rules will be regarded as significant just by random chance. Such rules are *false positives*. The number of rules being tested in an association rule mining task often reaches tens of thousands or even more. It is thus necessary to adjust the cut-off p-value threshold to reduce false positives. Some readers may argue that we can use $min\_sup$ and $min\_conf$ to eliminate false rules. The problem is that it is often very difficult for users to decide proper values for the two thresholds. If the two thresholds are low, then they cannot remove all false rules; if they are set to be high, then we are running the risk of throwing many real rules away. Thus, we cannot depend on the two thresholds alone to remove false rules.

An association rule is a testing of the association between its two sides, so association rule mining is a multiple testing problem. Several multiple testing correction methods have been proposed to control false positives in statistics. However, there is a lack of comprehensive study on their performance for the association rule mining task. In this paper, we conduct extensive experiments to study the ability of these methods in controlling false positives and in detecting real association rules under different settings.

The rest of the paper is organized as follows. Section 2 defines the problem. Section 3 briefly describes the association rule mining algorithm. The three multiple testing correction approaches are presented in Section 4. We describe experiment design and report experiment results in Section 5. Related work is described in Section 6. Finally, Section 7 concludes the paper.

## 2. PROBLEM DEFINITION

We consider a special type of association rules—class association rules [11] or predictive association rules [13, 21], which have been used for classification successfully. The definitions and methods described in the paper can be easily extended to other forms of association rules.

### 2.1 Class association rule

Class association rules are generated from attribute-valued data with class labels. Only class labels are allowed on the right-hand side. Let $D = \{t_1, t_2, \cdots, t_n\}$ be a set of records. Each record is described by a set of attributes $\mathcal{A} = \{A_1, A_2, \cdots, A_m\}$ and a class label attribute $C$. We assume all the attributes are categorical. If there are continuous attributes, we can discretize them using a supervised discretization method.

Let $A$ be an attribute and $v$ be a value taken by $A$. We call attribute-value pair $A = v$ an *item*. If an attribute $A$ of a record $t$ takes value $v$, then we say $t$ *contains* item $A = v$. We use letter $i$ to denote items.

DEFINITION 1 (PATTERN). *A pattern is a set of items $\{i_1, i_2, \cdots, i_k\}$, and $k$ is called the length of the pattern.*

We use letter $X$ to denote patterns. Given two patterns $X_1$ and $X_2$, if every item of $X_1$ is also contained in $X_2$, then $X_1$ is called a *sub-pattern* of $X_2$ and $X_2$ is called a *super-pattern* of $X_1$, denotated as $X_1 \subseteq X_2$ or $X_2 \supseteq X_1$. If a record $t$ contains all the items in a pattern $X$, then we say $t$ *contains* $X$, denoted as $X \subseteq t$ or $t \supseteq X$. The *support* of a pattern $X$ in a dataset $D$ is defined as the number of records in $D$ containing $X$. That is, $supp(X) = |\{t|t \in D \wedge X \subseteq t\}|$.

DEFINITION 2 (ASSOCIATION RULE). *An association rule takes the form: $X \Rightarrow c$, where $X$ is a pattern and $c$ is a class label.*

We use letter $R$ to denote rules. Given a rule $R : X \Rightarrow c$, if a record $t$ contains $X$, and its class label is $c$, then we say $t$ *supports* $R$. The *support* of a rule $R$ in a dataset $D$ is defined as the number of records in $D$ that support $R$, denoted as $supp(R)$. The *confidence* of $R$ is defined as the proportion of records labeled with class $c$ among those records containing $X$. That is, $conf(R) = supp(R)/supp(X)$. The support of $X$ is called the *coverage* of $R$.

Given a dataset $D$, a minimum support threshold $min\_sup$ and a minimum confidence threshold $min\_conf$, the association rule mining task aims to find all the rules $R : X \Rightarrow c$ such that $supp(X) \geq min\_sup$ and $conf(R) \geq min\_conf$. If the coverage of a rule is no less than $min\_sup$, then we say the rule is *frequent*.

### 2.2 P-value of class association rules

The p-value of rule $R : X \Rightarrow c$ is the probability of observing $R$ or a rule more extreme than $R$ if $X$ and $c$ are independent. Several statistical tests have been used to calculate p-values of association rules, like $\chi^2$ test [5] and Fisher's exact test [18, 19]. Here we adopt two-tailed Fisher's exact test [8] to calculate the p-value of a rule $R : X \Rightarrow c$ as follows:

$$\begin{aligned} p(R) &= p(supp(R); n, n_c, supp(X)) \\ &= \sum_{k \in E} H(k; n, n_c, supp(X)) \\ &= \sum_{k \in E} \frac{\binom{n_c}{k} \cdot \binom{n-n_c}{supp(X)-k}}{\binom{n}{supp(X)}} \end{aligned}$$

where $n$ is the total number of records in the given dataset, $n_c$ is the number of records labeled with class $c$, $H(k; n, supp(X), n_c)$ is the hypergeometric distribution, $\binom{a}{b}$ is binomial coefficient and $E$ is the set of cases that are equally extreme as $R$ or are more extreme than $R$, that is, $E = \{k|H(k; n, n_c, supp(X)) \leq H(supp(R); n, n_c, supp(X))\}$.

The p-value of a rule measures the statistical significance of the rule. If a rule $R$ has low p-value, it means that $R$ is unlikely to occur if $X$ and $c$ are independent. Given that $R$ occurs in the data, then $X$ and $c$ are unlikely to be independent, that is, $X$ and $c$ are likely to be associated. The lower the p-value, the more statistically significant the rule is.

Given a dataset $D$, the number of records in $D$ and the number of records labeled with class $c$ in $D$ are fixed. The p-value of a rule is decided by its coverage and confidence. The higher the coverage and the confidence, the lower the p-value as shown in Figure 1.

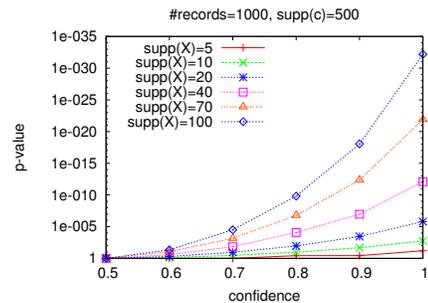

**Figure 1: p-values of rule $R : X \Rightarrow c$ under different $supp(X)$ and $conf(R)$. #records=1000, $supp(c)$=500.**

Note that here we are interested in finding the association between a pattern and a class label. We are not interested in the association between items within a pattern. If users are interested in the latter or other aspects of association rules, they may need to use other statistical tests to calculate p-values. The multiple testing correction approaches discussed in Section 4 can be applied as well.

### 2.3 Controlling false positives

When one single rule is tested, a p-value of 0.05 is often used as a cut-off threshold to decide whether a rule is *statistically significant* [6]. A p-value threshold of 0.05 means that there is a 0.05 probability that a rule is not real but we are wrongly regarding it as real. Such rules are *false positives* or *false discoveries*. The cut-off p-value threshold reflects the level of false positive error rate that a user is willing to accept. In the case that many rules are tested, the number of rules that are wrongly regarded as significant can be large. We need to adjust the cut-off threshold to control false positive errors under certain level. False positives can be controlled based on two measures: family-wise error rate (FWER) and false discovery rate (FDR) [4].

DEFINITION 3 (FWER). *Family-wise error rate is the probability of reporting at least one false positive.*

DEFINITION 4 (FDR). *False discovery rate is the expected proportion of false positives among the rules that are reported to be statistically significant.*

Obviously, FWER is more stringent than FDR. For testing problems where the goal is to provide definitive results, FWER is preferred. If a study is viewed as exploratory, control of FDR is often preferred. FDR allows researchers to identify a set of "candidate positives" of which a high proportion are likely to be true. The true positives within the candidate set can then be identified in a follow-up study. Association rule mining is exploratory in nature, hence FDR is often preferred for association rule mining.

A rule with low coverage cannot be very significant. For example, when #records=1000, $supp(c)$=500 and $supp(X)$=5, even if $conf(R)$=1, the p-value of $R : X \Rightarrow c$ is as high as 0.062. The same is true for rules with low confidence. When #records=1000 and $supp(c)$=500 and $conf(R)$=0.55, even if $supp(X)$=200, the p-value of $R$ is as high as 0.133. Some readers may argue that we can use the minimum support threshold and the minimum confidence threshold to eliminate false positives. The problem is that association rules do not have the same level of coverage and confidence. For rules with moderate confidence, we may need to use a high $min\_sup$ threshold to ensure that they are statistically significant. For rules with moderate coverage, we may need to use a high $min\_conf$ threshold. If we set both thresholds unnecessarily high, then many real rules may be thrown away. Hence it is not practical to use the two thresholds alone to control false positives.

Note that though we emphasize the statistical significance of association rules, we do not claim that p-value should replace confidence or many other interestingness measures proposed in the literature. The main role of the minimum confidence threshold is to reflect the level of domain significance. It answers the question "what is the minimum level of confidence that can be considered as interesting in this domain?". The level of domain significance is independent of sample size, and it should be decided by only domain experts. We believe that statistical significance measures and domain significance measures should be used together to filter uninteresting rules from different perspectives.

## 3. CLASS ASSOCIATION RULE MINING

Many algorithms have been developed to mine frequent patterns or association rules. We map every attribute-value pair to an item, and use an existing frequent pattern mining algorithm [12] to mine frequent patterns with support no less than $min\_sup$. Besides counting the support of a pattern, we also count the frequency of the class labels in the set of records containing the pattern to calculate the confidence and p-value of the corresponding rules. When there are only two class labels, $c$ and $\bar{c}$, in a dataset, testing $X \Rightarrow c$ is equivalent to testing $X \Rightarrow \bar{c}$. Hence when there are two class labels, we generate one rule for each pattern. When there are more than two class labels, we generate $m$ rules for each pattern, where $m$ is the number of class labels.

Frequent patterns often contain a lot of redundancy. Different patterns may represent the same set of records. If two patterns, $X_1$ and $X_2$, appear in the same set of records, then $X_1 \Rightarrow c$ and $X_2 \Rightarrow c$ have the same coverage and confidence. Consequently, their p-values are the same too. To reduce the number of rules generated, we use only closed frequent patterns [14] as the left-hand side of rules. A closed frequent pattern is the longest pattern among those patterns that occur in the same set of records as it, and it is unique.

## 4. MULTIPLE TESTING CORRECTION

Several multiple testing correction methods have been proposed. We categorize these methods into three categories: the direct adjustment approach, the permutation-based approach and the holdout approach.

### 4.1 The direct adjustment approach

**Bonferroni correction** [1] is one of the most commonly used approaches for multiple testing. It aims at controlling FWER. To maintain FWER at $\alpha$, Bonferroni correction divides the $\alpha$ threshold by the total number of tests performed. Let $N_t$ be the number of tests performed, then those tests with p-value no larger than $\frac{\alpha}{N_t}$ are regarded as statistically significant and others are not.

In the class association rule mining task, the number of tests performed is $m \cdot N_{FP}$, where $N_{FP}$ is the number of patterns with support no less than $min\_sup$ and $m$ is the number of class labels if the number of class labels is larger than 2, $m = 1$ if the number of class labels is 2.

**Benjamini and Hochberg's method** [4] controls false positive rate (FDR). Let $H_1, H_2, \cdots, H_n$ be the $n$ tests and they are sorted in ascending order of p-value. Their corresponding p-values are $p_1, p_2, \cdots, p_n$. To control FDR at a level of $\alpha$, this method finds the largest $i$, denoted as $k$, for which $p_i \leq \frac{i \cdot \alpha}{n}$, and then regards all $H_i$, $i$=1, 2, $\cdots$, $k$, as statistically significant.

### 4.2 The permutation-based approach

The permutation-based approach [20, 7] randomly shuffles the class labels of the records and recalculate the p-value of the rules. The random shuffling destroys the association between patterns and class labels, hence the distribution of the re-calculated p-values is an approximation of the null distribution where the two sides of rules are independent.

To control FWER at a level of $\alpha$, we randomly generates $N$ permutations. There should be no real rules on a permutation, hence any rule this is declared to be statistically significant on a permutation is a false positive. We need to find a cut-off p-value threshold such that the proportion of permutations on which at least one rule passes the cut-off threshold is no larger than $\alpha$. To find this cut-off threshold, we get the lowest p-value on each permutation and rank them in ascending order. The $\lfloor \alpha \cdot N \rfloor$-th p-value is then used as the cut-off threshold to decide whether a rule is statistically significant.

To control FDR at a level of $\alpha$, we randomly generates $N$ permutations and adjusts the p-value of individual rules as follows. Let $N_t$ be the number of rules tested on the original dataset, $H = \{p_1, p_2, \cdots, p_{N \cdot N_t}\}$ be the p-values of the $N_t$ rules on the $N$ permutations and $p$ be the p-value of a rule $R$ on the original dataset. Then the new p-value of rule $R$ is re-calculated $\frac{|\{p_i | p_i \leq p, \ p_i \in H\}|}{N \cdot N_t}$. Benjamini and Hochberg's method is then applied on the new p-values to find the cut-off threshold.

The permutation-based approach preserves the interactions among patterns, so it can find a more accurate cut-off p-value threshold than the direct adjustment approach. However, the permutation-based approach is very costly. We use several techniques to reduce its cost.

### 4.2.1 Mining association rules only once

Association rule mining can be very costly, so it is not desirable to perform association rule mining on each permutation. Class labels of records change over different permutations, but other items in the records do not change. Given a rule $R : X \Rightarrow c$, $X$ occurs in the same set of records on all the permutations as on the original dataset, so $supp(X)$ does not change across different permutations, but $supp(R)$ changes due to the shuffling of class labels. We mine frequent patterns only once on the original dataset and generate the record id lists of frequent patterns. The supports and p-values of the rules on a permutation are calculated using the record id lists and the class labels of that permutation.

### 4.2.2 Diffsets

The record id lists of frequent patterns can be very long. To further reduce the cost, we use a technique called *Diffsets*. This technique was first proposed in [22] for improving the performance of a frequent pattern mining algorithm. Frequent patterns can be organized in a set-enumeration tree [16]. We use a depth-first order to explore the set-enumeration tree. The record id list of a pattern $X$ is generated from that of its parent in the tree. We denote the parent as $Y$. The basic idea of *Diffsets* is that if $supp(X)$ is very close to $supp(Y)$, then we can store the difference instead of the full record id list of $X$. More specifically, if $supp(X) <= supp(Y)/2$, then we store the full record id list of $X$; otherwise, we store the difference between the two record id lists, denoted as *Diffsets(X)*. That is, *Diffsets(X)* contain the ids of the records that contain $Y$ but does not contain $X$. If *Diffsets(X)* is stored, $supp(X \Rightarrow c)$ is calculated from $supp(Y \Rightarrow c)$ and *Diffsets(X)*.

### 4.2.3 Buffering p-values

Let $N_t$ be the number of rules tested on the original dataset and $N$ be the number of permutations. We need to calculate $N_t \cdot (N+1)$ p-values in the permutation-based approach. This can be very costly. Fortunately, the calculation of p-values can be shared between different rules and across different permutations. We store p-values that are previously computed to enable the sharing.

Let $n$ be the number of records in a given dataset. The calculation of $H(k; n, supp(c), supp(X))$ requires the factorials of several integers. To speed up the calculation, we store the factorials of the integers from 0 to $n$ in a memory buffer of size $n+1$. We denote this buffer as $B_f$. The $n+1$ factorials can be calculated incrementally in $O(n+1)$ time. If $n$ is large, the factorial of $n$ may exceed the range of the double data type. To solve this problem, we store the logarithm of the factorials in the buffer.

Given a rule $R : X \Rightarrow c$, we need to find the set of cases that are equally extreme as $R$ or are more extreme than $R$ to get the p-value of $R$. We proceed as follows. We get the lower bound $L$ and upper bound $U$ of $supp(R)$, where $L = \max\{0, supp(c) + supp(X) - n\}$ and $U = \min\{supp(c), supp(X)\}$. We compute $H(k; n, supp(c), supp(X))$ for all $k \in [L, U]$ using the factorials stored in $B_f$, and we store them in another memory buffer of size ($U$-$L$+1). We call this new buffer the p-value buffer of $supp(X)$ and denote it as $B_{supp(X)}$. Based on the property of the hypergeometric distribution, the most extreme cases are located on the two ends of $B_{supp(X)}$. In other words, $H(L; n, supp(c), supp(X))$ and $H(U; n, supp(c), supp(X))$ are the two smallest values in the buffer. When we move toward the middle of the buffer, $H(k; n, supp(c), supp(X))$ becomes larger and larger. Figure 2 shows the values stored in $B_{supp(X)}$ when $n=20$, $supp(c)=11$ and $supp(X)=6$.

Buffer B

| k | H(k; 20, 11, 6) | sum-up order |
|---|---|---|
| 0 | 0.0021672 | 0 |
| 1 | 0.035759 | 2 |
| 2 | 0.17879 | 4 |
| 3 | 0.35759 | 6 |
| 4 | 0.30650 | 5 |
| 5 | 0.10728 | 3 |
| 6 | 0.011920 | 1 |

Buffer B: after sum-up

| k | p(k; 20, 11, 6) | |
|---|---|---|
| 0 | 0.0021672 | =H(0; 20,11, 6) |
| 1 | 0.049845 | =p(6; 20, 11, 6)+H(1; 20, 11, 6) |
| 2 | 0.33591 | =p(5; 20, 11, 6)+H(2; 20, 11, 6) |
| 3 | 1.0000 | =p(4; 20, 11, 6)+H(3; 20, 11, 6) |
| 4 | 0.64241 | =p(2; 20, 11, 6)+H(4, 20, 11, 6) |
| 5 | 0.15712 | =p(1; 20, 11, 6)+H(5; 20, 11, 6) |
| 6 | 0.014087 | =p(0; 20, 11, 6)+H(6; 20, 11, 6) |

**Figure 2: An example of p-value buffer $B_{supp(X)}$ and its calculation.** $n$=20, $supp(c)$=11, $supp(X)$=6.

To get all the possible p-values that a rule with coverage $supp(X)$ can have, we start from the two ends of the buffer and move towards the middle, and sum up the values one at a time in ascending order of $H(k; n, supp(c), supp(X))$. Let $p$ be the sum. Initially, $p$=0. Let $H(k; n, supp(c), supp(X))$ be the next value to be added to $p$, then $p + H(k; n, supp(c), supp(X))$ is the p-value of $R$ when $supp(R) = k$. We use $p+ H(k; n, supp(c), supp(X))$ to replace $H(k; n, supp(c), supp(X))$ in the buffer. When all $H(k; n, supp(c), supp(X))$ are summed up, where $k \in [L, U]$, buffer $B_{supp(X)}$ stores all the possible p-values that a rule with coverage $supp(X)$ can have. The calculation is illustrated in Figure 2. The time complexity for calculating the values in $B_{supp(X)}$ is $O(U - L + 1)$.

The coverage of a rule does not change over different permutations, only its support changes. Therefore, given a rule $R : X \Rightarrow c$, we need to calculate $B_{supp(X)}$ only once. The p-values of $R$ on the $N$ permutations can be retrieved directly from the buffer.

Different rules may have the same coverage, and the computation of their p-values can be shared too. To enable the sharing between different rules, we use a static buffer and a dynamic buffer. The static buffer stores the p-value buffers of the rules with coverage between $min\_sup$ and $max\_sup$. The value of $max\_sup$ is decided by the size of the static buffer. If the coverage of a rule is larger than $max\_sup$, then the p-value buffer of the rule is stored in the dynamic buffer. The dynamic buffer is much smaller than the static buffer, and its contents are updated constantly. The dynamic buffer always stores only one p-value buffer, which is the p-value buffer of the last rule whose coverage is larger than $max\_sup$. We use a variable $sup_d$ to remember whose p-value buffer the dynamic buffer is storing.

When we calculate the p-value of a rule $R : X \Rightarrow c$, we first check whether $supp(X) \leq max\_sup$. If it is true, then we look for the p-value buffer $B_{supp(X)}$ in the static buffer. If $B_{supp(X)}$ has not been calculated before, we calculate it and store it in the static buffer. The p-value of $R$ is then retrieved from there. If $supp(X) > max\_sup$, we check whether $supp(X) = sup_d$. If it is true, then we retrieve the p-value of $R$ directly from the dynamic buffer. Otherwise, we calculate the p-values in $B_{supp(X)}$, and store them in the dynamic buffer. The value of $sup_d$ is set to $supp(X)$. We then get the p-value of $R$ from the dynamic buffer.

### 4.3 The holdout approach

The holdout evaluation approach is proposed by Webb [18]. It aims to overcome the drawbacks of the above two approaches. It divides a dataset into an exploratory dataset and an evaluation dataset. Association rules are first mined from the exploratory dataset. The set of rules with p-value no greater than $\alpha$ are then passed to the evaluation dataset for validation. To control FWER at level $\alpha$, the p-value of the rules on the evaluation dataset is adjusted using Bonferroni correction, but now the number of tests is the number of rules that have a p-value no larger than $\alpha$ on the exploratory dataset. Typically, that number is orders of magnitude smaller than the number of rules being tested on the whole dataset, thus the holdout approach is expected to have a better chance of discovering rules with a moderately low p-value. FDR is controlled in a similar way using Benjamini and Hochberg's method.

The holdout approach is less costly than the permutation-based approach. However, the performance of the holdout approach may be affected by the way the dataset is partitioned. If a rule happens to fall in only the exploratory dataset or the evaluation dataset, then this rule cannot be discovered. The coverage of the rules on the exploratory dataset and the evaluation dataset is almost halved, so rules have much higher p-values on the exploratory dataset and the evaluation dataset. This on one hand makes some true association rules undetectable, on the other hand, it becomes harder for noise rules to turn out significant.

## 5. A PERFORMANCE STUDY

In this section, we study the performance of the three multiple correction approaches. Our experiments were conducted on a PC with a 2.33Ghz CPU and 4GB memory.

### 5.1 Datasets

It is very hard to know the complete set of true association rules in real-world datasets, so it is difficult to evaluate the performance of the three approaches on real-world datasets. To solve this problem, we generate synthetic datasets and embed rules in them. We generate synthetic datasets in matrix forms, where rows represent records and columns represent attributes. All the attributes are categorical. We first embed a number of association rules in the matrix. The cells that are not covered by any embedded rules are then filled randomly. If no rule is embedded, then the data is totally random. The parameters taken by the data generator are listed in Table 1.

| | |
|---|---|
| $N$ | number of records |
| $\#C$ | number of classes |
| $A$ | number of attributes |
| $min\_v, max\_v$ | Minimum and maximum number of values taken by an attribute |
| $N_r$ | #rules embedded |
| $min\_l, max\_l$ | Minimum and maximum length of embedded rules |
| $min\_s, max\_s$ | Minimum and maximum coverage of embedded rules |
| $min\_c, max\_c$ | Minimum and maximum confidence of embedded rules |

**Table 1: Parameters used by the synthetic data generator**

For the experiments below, some parameters of the synthetic dataset generator are fixed to the following values: $\#C=2$, $min\_v=2$, $max\_v=8$, $min\_l=2$ and $max\_l=16$. The records are evenly distributed in different classes. We have tried other parameter settings, like setting the number of classes $\#C$ to be larger than 2. The results we obtained are similar to the results reported below.

The performance of the holdout approach may be affected by the way the dataset is partitioned. To have a fair comparison of the holdout approach, we generate two sub-datasets with $N/2$ records and embed rules with coverage between $min\_s/2$ and $max\_s/2$ into them. We then catenate the two sub-datasets into a single dataset with $N$ records and the embedded rules in this dataset will have coverage between $min\_s$ and $max\_s$. For the holdout evaluation, we use one of the two sub-datasets as the exploratory dataset, and the other one as the evaluation dataset. This way, the impact of the partitioning is eliminated. We call this method "holdout". We also tried random partitioning in our experiments, and we call it "random holdout". In all the experiments, the minimum support threshold $min\_sup$ on the exploratory dataset is set to be half of that on the whole dataset.

| Datasets | #records | #attributes | #classes |
|---|---|---|---|
| adult | 32561 | 14 | 2 |
| german | 1000 | 20 | 2 |
| hypo | 3163 | 25 | 2 |
| mushroom | 8124 | 22 | 2 |

**Table 2: Real-world datasets**

Besides synthetic datasets, we also used four real-world datasets downloaded from UCI machine learning repository[1] in our experiments. The four datasets are listed in Table 2. Continuous attributes in these datasets are discretized using MLC++ [2].

---

[1] http://archive.ics.uci.edu/ml/
[2] http://www.sgi.com/tech/mlc/

## 5.2 Evaluation method

If we embed one rule $X \Rightarrow c$ in a synthetic dataset, then the sub-patterns and super-patterns of $X$ are likely to form significant association rules with $c$ too. Figure 3 shows the distribution of p-values on three datasets: a random dataset without embedded rules, two datasets with one embedded rule. The coverage of the two embedded rules is set to 400 and 200 respectively and their confidence is set to 0.8. For all the three datasets, $N=2000$ and $A=40$. Figure 3 shows that one embedded rule leads to many other rules with low p-values. These by-product rules should not be simply treated as false positives. Otherwise, the FDR of all the correction methods will be close to 1.

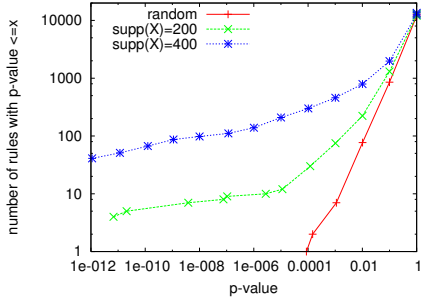

**Figure 3:** Distribution of p-values in three cases. $N$=**2000**, $A$=**40**, $conf(R)$=**0.8**.

When we embed only one rule $R_t : X_t \Rightarrow c_t$ in a synthetic dataset, we define false positive as follows. Let $\alpha$ be the cut-off p-value threshold. Let $T(X)$ be the set of records containing pattern $X$. A rule $R : X \Rightarrow c$ with p-value no larger than $\alpha$ is called a *false positive* if $R \neq R_t$ and $R$ satisfies one of the following conditions:

- $T(X_t) \bigcap T(X) = \phi$;
- $T(X_t) \bigcap T(X)$ is not empty, and $p(R|\neg R_t) \leq \alpha$, where $p(R|\neg R_t)$ is the adjusted p-value of $R$ if $R_t$ does not exist.

The definition of $p(R|\neg R_t)$ is given below. Let $n$ be the number of records in the given dataset $D$ and $n_{c_t}$ be the support of $c_t$ in $D$. If $R_t$ does not exist, then the proportion of $c_t$ in $T(X_t) \bigcap T(X)$ should be close to $\frac{n_{c_t}}{n}$. We use $supp(X \cup X_t) \cdot \frac{n_{c_t}}{n}$ to approximate the expected support of $c_t$ in $T(X) \bigcap T(X_t)$ if $R_t$ does not exist. The adjusted support of $R$ on the whole dataset, if $R_t$ does not exist, can then be calculated as $supp(R|\neg R_t) = supp(X \cup X_t) \cdot \frac{n_{c_t}}{n} + (supp(R) - supp(X \cup X_t \cup c))$. The adjusted p-value of $R$ if $R_t$ does not exist is defined as $p(R|\neg R_t) = p(supp(R|\neg R_t); n, n_c, supp(X))$.

Based on the above definition of false positive, we define power, FWER and FDR accordingly. On a single dataset,

- FWER is 1 if there is at least one false positive; otherwise FWER is 0.
- FDR is the proportion of false positives among all the rules that are reported to be statistically significant.
- power is the proportion of the embedded rules that are reported to be statistically significant. When only one rule is embedded, power is either 1 or 0.

In our experiments, we generate 100 datasets for each parameter setting of the synthetic dataset generator, and report the average results on the 100 datasets. On these 100 datasets,

- FWER is defined as the proportion of datasets that have at least one false positive.
- FDR is the average FDR over the 100 datasets.
- power is the average power over the 100 datasets. If only one rule is embedded, power is also the proportion of the datasets on which the embedded rule is detected.

The results reported below were obtained by controlling FWER and FDR at 5%. We have tried to control FWER and FDR at other levels, like 1% and 0.1%. At these two error levels, all the three approaches have lower power and lower error rate than that at 5%, but their relative performance is the same as that at 5%. In all the experiments, we set the minimum confidence threshold to 0.

## 5.3 Running time

The first experiment compares the running time of the three correction approaches. The four real-world datasets listed in Table 2 and two synthetic datasets are used in this experiment. Dataset D8hA20R0 is generated using the following parameters: $N=$ 800, $A=20$ and $N_r=0$. Dataset D2kA20R5 is generated using the following parameters: $N=$ 2000, $A=20$, $N_r=5$, $min\_s=400$, $max\_s=600$, $min\_c=0.6$ and $max\_c=0.8$. In all the experiments, the number of permutations is set to 1000.

We first study how much the *Diffsets* technique and the p-value buffering technique described in Section 4.2 improve the efficiency of the permutation-based approach. Figure 4 shows the running time of the permutation-based approach in four cases: (1) association rules are mined only once, but the *Diffsets* technique and the p-value buffering technique are not used, denoted as "no optimization"; (2) only the dynamic buffer is used, denoted as "dynamic buffer"; (3) the dynamic buffer and the *Diffsets* technique is used, denoted as "*Diffsets*+dynamic buffer"; (4) a 16MB static buffer is used in addition to *Diffsets* and the dynamic buffer, denoted as "16M static buffer+*Diffsets*+dynamic buffer". In all the figures, the running time includes frequent pattern mining time and multiple testing correction time.

Using the dynamic buffer to store pre-computed p-values can speed-up permutation test by an order of magnitude on almost all the datasets. The *Diffsets* technique further reduces the running time by 2 to 10 times on the four largest datasets. On the random dataset D8hA20R0, the size of the *Diffset* of a pattern is very close to that of the full record id list of the pattern, hence *Diffsets* cannot achieve any improvement. The static buffer does not achieve further improvement given the dynamic buffer has already been used.

Figure 5 shows the running time of the three correction approaches. The permutation-based approach uses the *Diffsets* technique and the p-value buffering technique. The direct adjustment approach incurs the lowest overhead. The permutation-based approach has the highest computation cost. It can be tens of times slower than the direct adjustment approach. The holdout approach is several times slower than the direct adjustment approach.

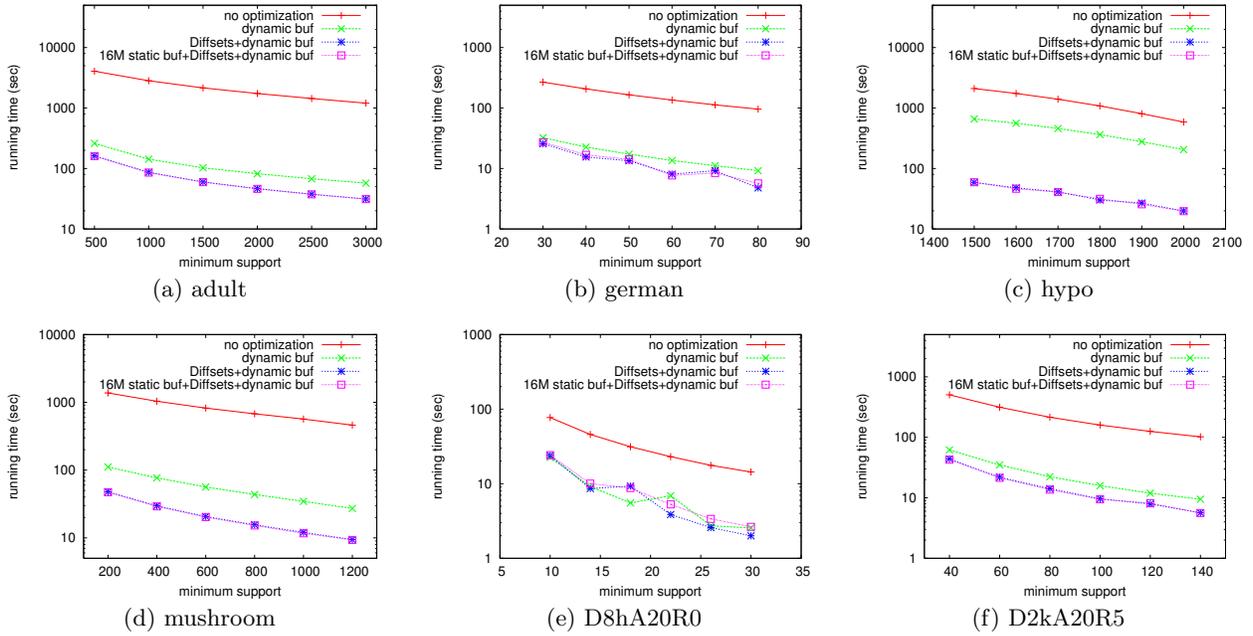

Figure 4: Improvements of the *Diffsets* technique and the p-value buffering technique.

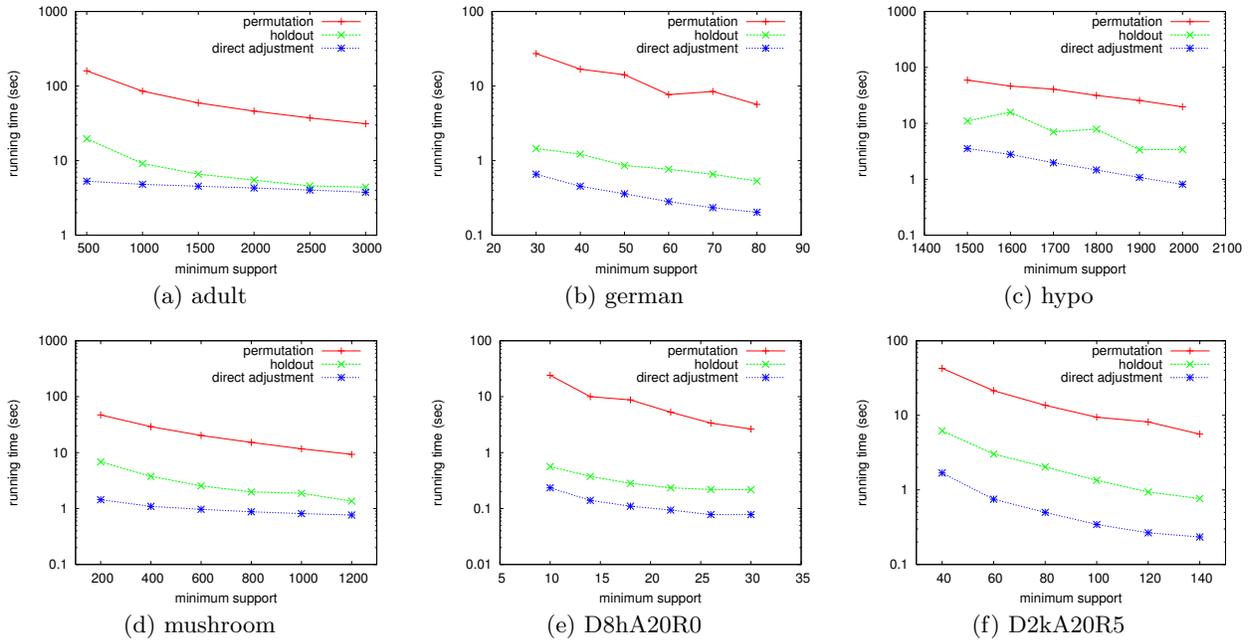

Figure 5: Running time of the three correction approaches.

## 5.4 Random datasets

The second experiment studies the ability of different approaches in controlling FWER and FDR. It is conducted on random datasets without embedding any rules, so every rule that is reported to be statistically significant is a false positive. FWER and average FDR over 100 datasets have the same meaning as FDR is either 0 or 1 on a random dataset. The random datasets are generated using the following parameters: $N=2000$, $A=40$ and $N_r=0$.

Figure 6 shows the performance of the three approaches when the minimum support threshold $min\_sup$ is varied. The meaning of the abbreviations in the figures are listed in Table 3. When $min\_sup$ decreases from 1000 to 100, the number of rules tested increases quickly as shown in Figure 6(b). The same trend is observed for FWER and the number of false positives when no correction is made. In particular, when $min\_sup \leq 200$, FWER reaches 1 if no correction is made. All the three correction approaches can control FWER at around 5%. The direct adjustment approach and the permutation-based approach have similar performance. The holdout approach has the lowest FWER, and it also produces the fewest number of false positives.

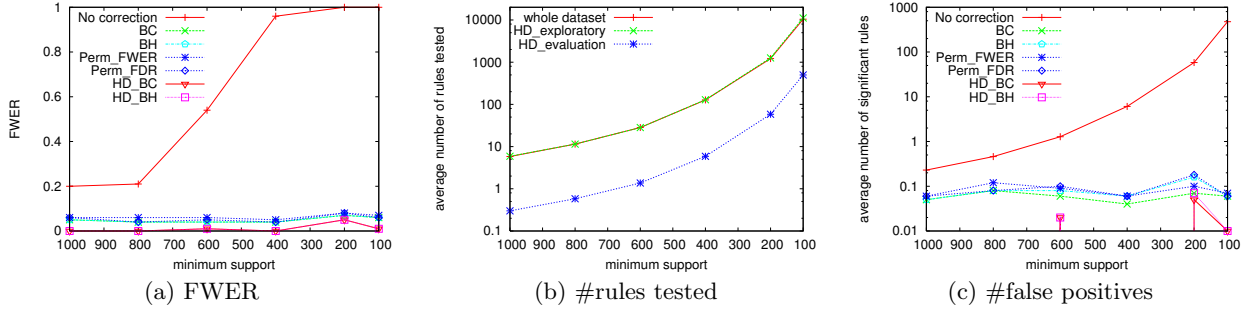

(a) FWER　　(b) #rules tested　　(c) #false positives

Figure 6: Performance of the three approaches on random datasets ($N$=2000, $A$=40). The meaning of the abbreviations can be found in Table 3

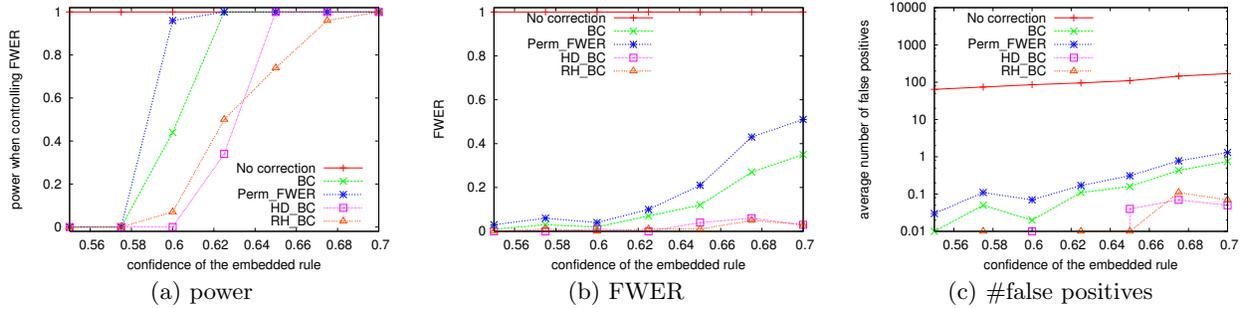

(a) power　　(b) FWER　　(c) #false positives

Figure 8: Performance of the three approaches on datasets with one embedded rule when FWER is controlled at 5%. $min\_sup$=150 on the whole dataset.

| Abbrv | Description |
|---|---|
| BC | Bonferroni correction |
| BH | Benjamini and Hochberg's method |
| Perm_FWER | Controlling FWER using permutation test |
| Perm_FDR | Controlling FDR using permutation test |
| HD | The holdout method on two sub-datasets |
| HD_BC | Holdout with Bonferroni correction |
| HD_BH | Holdout with Benjamini and Hochberg's method |
| RH | The holdout method using random partitioning |
| RH_BC | Random holdout with Bonferroni correction |
| RH_BH | Random holdout with Benjamini and Hochberg's method |

Table 3: Abbreviations

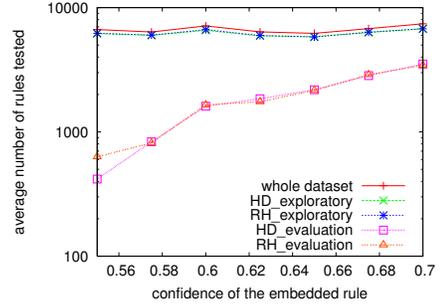

Figure 7: Number of rules tested.

## 5.5 Datasets with one rule embedded

This experiment studies the power of the three approaches in detecting embedded rules. We embed only one rule in each dataset, and we use $R_t : X_t \Rightarrow c_t$ to denote the embedded rule. We generate 100 datasets using the following parameters: $N$=2000, $A$=40, $N_r$=1, $min\_s$=$max\_s$ =400. The confidence of $R_t$ is varied from 0.55 to 0.70.

### 5.5.1 Controlling FWER at 5%

Figure 8 shows the performance of the three approaches when FWER is controlled at 5%. Figure 7 shows the number of rules tested. The embedded rule can always be detected when no correction is made, but it is at the cost of high FWER as shown in Figure 8(b). When no correction is made, FWER is always 1 and the number of false positives is also considerably large as shown in Figure 8(c). The power of the three correction approaches increases when $conf(R_t)$ increases. In particular, when $conf(R_t)$ =0.55, none of the correction approaches can detect the embedded rule; when $conf(R_t)$=0.7, all the correction approaches can detect the embedded rule. This is because p-value decreases dramatically when $conf(R_t)$ increases as shown in Figure 9, which makes $R_t$ much easier to detect.

The permutation-based approach has higher power than the direct adjustment approach. When $conf(R_t)$=0.6, the permutation-based approach can detect the embedded rule on almost all the datasets, so its power is close to 1. The direct adjustment approach can detect the embedded rule on only 44 datasets out of the 100 datasets. It indicates that the cut-off p-value threshold decided by the direct adjustment approach is too low, which introduces many false negatives. The holdout approach has lower power than the other two approaches. The low power of the holdout approach is attributed to the fact that the p-value of $R_t$ is very sensitive to the coverage of $R_t$. On both the exploratory dataset and the evaluation dataset, the coverage of $R_t$ is reduced to half and its p-value is increased by several orders as shown in Figure 9, which makes $R_t$ undetectable in some cases.

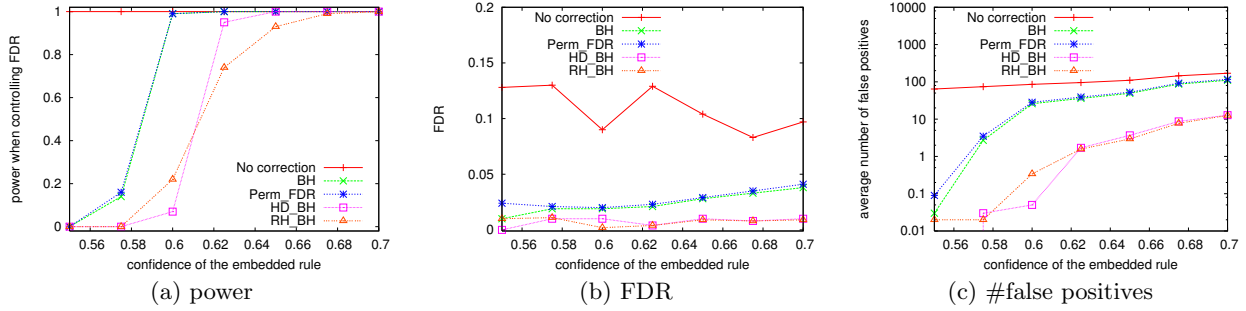

(a) power           (b) FDR           (c) #false positives

**Figure 10**: Performance of the three approaches on datasets with one embedded rule when FDR is controlled at 5%. $min\_sup$=150 on the whole dataset. The meaning of the abbreviations can be found in Table 3.

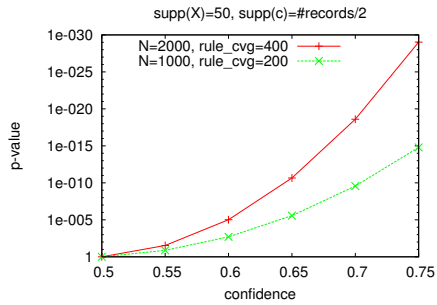

**Figure 9**: p-values under different $N$, $coverage(R_t)$ ($rule\_cvg$) and $conf(R_t)$ (the X-axis). $N_c$=$N$/2.

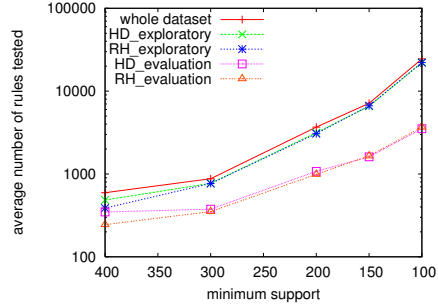

**Figure 11**: Number of rules tested under different $min\_sup$. $conf(R_t)$=**0.60**.

When $conf(R_t)$ increases, the holdout approach maintains low FWER, while the FWER of the other two approaches increases. When $conf(R_t)$=0.7, the FWER of the permutation-based approach even reaches 50%, which is much larger than the expected value of 5%. One possible reason is that when we embed a rule $R_t : X_t \Rightarrow c_t$ in a dataset, not only the class distribution in the set of records containing $X_t$ is distorted, the class distribution in the other part of the data is distorted too. The latter distortion can also produce some rules with low p-values and they are regarded as false positives. If we look at the absolute number of false positives generated by the permutation-based approach, it remains very low as shown in Figure 8(c). It is around 1 when $conf(R_t)$=0.7.

### 5.5.2 Controlling FDR at 5%

Figure 10 shows the performance of the three approaches when FDR is controlled at 5%. Again, the holdout approach has the lowest power, the lowest FDR and the fewest number of false positives. The direct adjustment approach and the permutation-based approach have very similar performance.

### 5.5.3 Impact of the number of rules tested

This experiment studies the impact of the number of rules tested on the performance of the several correction methods. We fix $conf(R_t)$ at 0.60, and vary the minimum support threshold to change the number of rules tested. Figure 11 shows the number of rules tested under different $min\_sup$. The X-axis is the minimum support threshold on the whole dataset. On the exploratory dataset, $min\_sup$ is set to be half of that on the whole dataset.

Figure 12 and Figure 13 shows the performance of the three correction approaches when the number of rules tested changes. When $min\_sup$ decreases, the number of rules tested increases. The three correction approaches need to use a lower cut-off p-value threshold to control false positives, which makes the embedded rule become undetectable sometimes, so the power of the three correction approaches decreases. The direct adjustment approach suffers a larger and faster drop in power than the permutation-based approach. When no correction is made, FWER and FDR increase slightly. For the three correction approaches, FWER and FDR decrease slightly, which indicates the three approaches are very effective at controlling false positives.

When $min\_sup$=400, the random holdout method has lower power than when $min\_sup$=300. The reason being that the coverage of the embedded rule $R_t$ is 400. When the random holdout approach divides the dataset into the exploratory dataset and the evaluation dataset randomly, the coverage of $R_t$ may be below 200 on the exploratory dataset, so it cannot be detected when $min\_sup$ is set to 200. Such cases are avoided when $min\_sup$ is lowered.

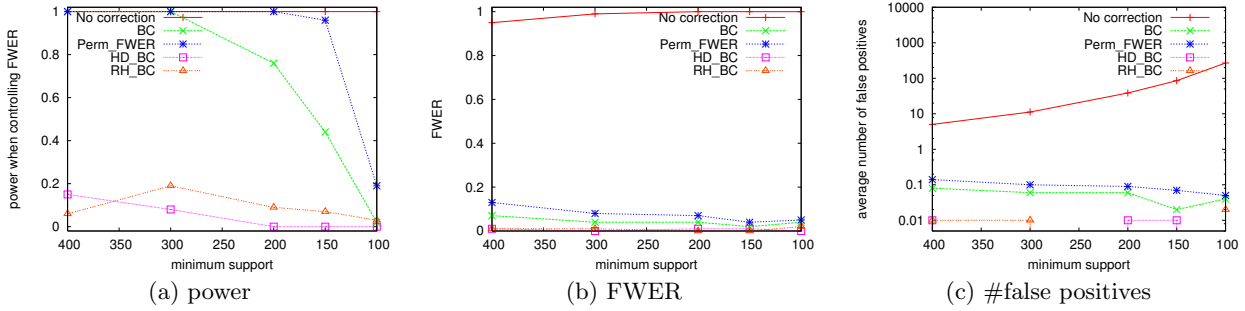

Figure 12: Impact of the number of rules tested when FWER is controlled at 5%. $conf(R_t)=0.60$.

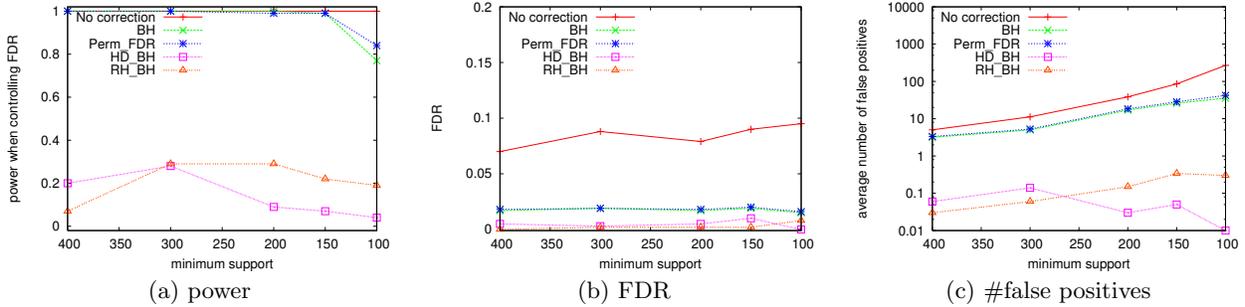

Figure 13: Impact of the number of rules tested when FDR is controlled at 5%. $conf(R_t)=0.60$. The meaning of the abbreviations can be found in Table 3.

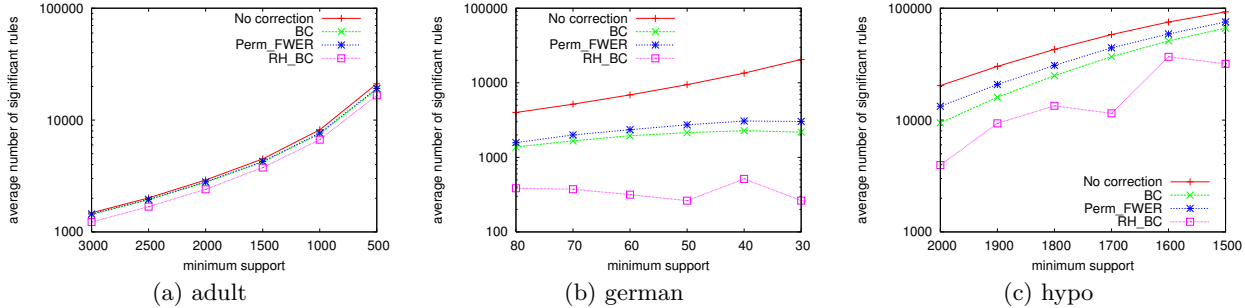

Figure 14: Number of significant rules reported on real-world datasets when FWER is controlled at 5%.

### 5.6 Results on real-world datasets

On real-world datasets, we cannot calculate power, FWER and FDR because real association rules are unknown. Here we compare the relative power and error rate of the three approaches by showing the number of significant rules reported by them. Approaches reporting more significant rules usually have higher power and higher error rate.

Figure 14 shows the number of significant rules when FWER is controlled at 5%. On *adult*, the three approaches produce a similar number of significant rules. The same is observed on *mushroom*. On the other two datasets, the permutation-based approach reports more significant rules than the direct adjustment approach, and both approaches produce much more rules than the holdout approach, which is consistent with the results on synthetic datasets.

The above results can be explained by Figure 15. On *adult* and *mushroom*, the p-value of more than 80% of the rules is below $10^{-12}$. These rules are reported to be significant by all the three approaches. On *hypo*, more than 30% of the rules have a p-value between $10^{-6}$ and $10^{-2}$. These rules are reported to be significant when no correction is made. The permuation-based approach regards about half of them as significant. The direct adjustment approach and the holdout approach regard none of them as significant. The situation is similar on dataset *german*.

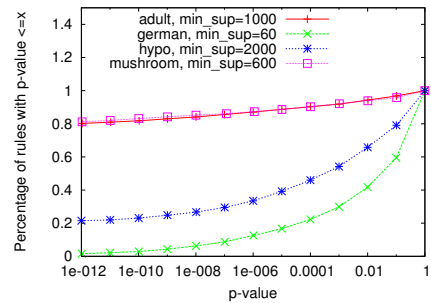

Figure 15: Distribution of p-values on real-world datasets

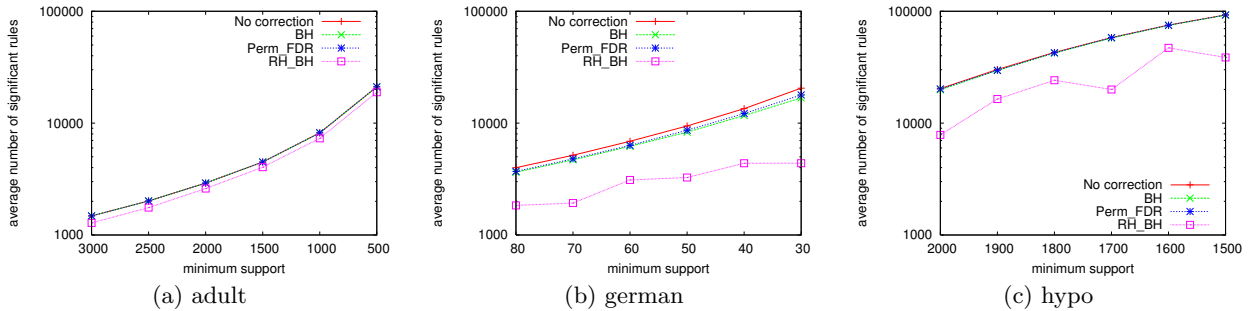

(a) adult     (b) german     (c) hypo

**Figure 16: Number of significant rules reported on real-world datasets when FDR is controlled at 5%.**

Figure 14 shows the number of significant rules when FDR is controlled at 5%. The number of significant rules reported by the direct adjustment approach and the permuation-based approach is very similar on all the four datasets. The holdout approach reports much fewer significant rules on *hypo* and *german*.

| p-value / conf | [0.75, 0.85) | [0.85, 0.9) | [0.9, 0.95) | [0.95, 1] |
|---|---|---|---|---|
| (0.05, 1] | 1586 | 0 | 0 | 0 |
| (0.01, 0.05] | 859 | 0 | 0 | 0 |
| (0.001, 0.01] | 724 | 323 | 0 | 0 |
| $(10^{-4}, 0.001]$ | 228 | 479 | 32 | 0 |
| $(10^{-5}, 10^{-4}]$ | 94 | 241 | 200 | 0 |
| $(10^{-6}, 10^{-5}]$ | 46 | 119 | 256 | 12 |
| $(10^{-7}, 10^{-6}]$ | 30 | 82 | 220 | 87 |
| $(10^{-8}, 10^{-7}]$ | 11 | 31 | 124 | 86 |
| $(0, 10^{-8}]$ | 16 | 77 | 279 | 289 |

**Table 4: Number of rules with different levels of confidence and p-value on dataset *german*. $min\_sup$=60.**

We use dataset *german* to show why it is difficult to use the minimum confidence threshold to eliminate statistically insignificant rules. Table 4 shows the number of rules with different levels of confidence and p-value on dataset *german*. The RHS of the rules is "class=good", and 70% of the records on the whole dataset have class label "good". The minimum support threshold is set to 60. The total number of rules tested is 13064. When FWER is controlled at 0.05, the cut-off p-value threshold decided by the direct adjustment approach and the permutation-based approach is $3.83 \times 10^{-6}$ and $1.83 \times 10^{-5}$ respectively. If we set $min\_conf$=0.85, then 834 (=323+429+32) of the reported rules have a p-value larger than $1 \times 10^{-4}$. They are not statistically significant according to the multiple testing correction approaches. If we increase $min\_conf$ to 0.9, then 247 (=30+11+16+82+31+77) rules with p-value lower than $1 \times 10^{-6}$ are discarded. These rules may represent real systematic effects. Hence using $min\_conf$ to eliminate insignificant rules may force us to use an unnecessarily high value for $min\_conf$, which may throw away many rules that are potentially real.

## 6. RELATED WORK

Since the association rule mining problem was first proposed by Agrawal *et al.*[2] in 1993, it has become an important problem in the data mining area. Association rule mining algorithms often produce a large number of rules. Various interestingness measures have been proposed to select rules. Tan *et al.*[17] and Geng *et al.*[9] surveyed various measures proposed in the literature. Many of the measures are defined based on support and confidence of rules, and they reflect domain significance of rules instead of statistical significance of rules.

There are a few papers studying the statistical significance of frequent patterns and association rules. Brin et al. [5] use the $\chi^2$ test to assess the statistical significance of individual rules, but they did not consider the effect of the number of rules being tested. Kirsch et al. [10] study the statistical significance of the frequency of frequent patterns instead of the association between the two sides of rules. They propose an algorithm to identify a threshold $s^*$ such that the set of patterns with support at least $s^*$ can be flagged as statistically significant with a small false discovery rate. Megiddo and Srikant [13] also study the statistical significance of the frequency of frequent patterns. They use re-sampling techniques to determine a proper p-value threshold. The samples are generated by preserving the frequency of single items, but the occurrences of all the items are independent. The p-values on these random datasets are used to determine the cut-off p-value threshold on the original dataset. However, the number of random datasets generated is 9, which may be too small to find a proper cut-off p-value threshold. Bay and Pazzani [3] use a Bonferroni-like correction to control false positives in contrast set mining.

Recently, Webb [18] investigates two methods to controlling false positives in association rule mining: the Bonferroni correction method [1] and the holdout evaluation approach. The p-value of a rule is calculated based on the rule's immediate subsets using Fisher's exact test, thus a p-value reflects the relationship between a rule and its subsets instead of the association between its two sides as in this paper. Webb later proposed another approach which uses layered critical values to control false positives [19]. The layered critical values are calculated based on the length of the rules. The above three methods are evaluated on datasets with a small number of items where the search space is small and the schema of the datasets is fixed. In this paper, we conducted a more comprehensive study to get a thorough understanding of different correction approaches.

## 7. DISCUSSION AND CONCLUSION

In this paper, we studied three multiple testing correction approaches for controlling false positives in association rule mining. Our findings can be summarized below.

- In terms of power, the order of the three approaches is permutation test > direct adjustment > holdout. In terms of error rate, the order is the same.

- In terms of computation cost, the order is permutation test > holdout > direct adjustment.

- The permutation-based approach has very close performance to the direct adjustment approach when FDR is controlled. Since the permutation-based approach is much more costly, the direct adjustment approach is more favorable when users want to control FDR.

- When FWER is controlled at $\alpha$ and a very small portion of rules have a p-value between $\frac{\alpha}{N_t}$ and $\alpha$, where $N_t$ is the number of rules tested, then it is not worthwhile to use the permutation-based approach. If many rules have a p-value between $\frac{\alpha}{N_t}$ and $\alpha$, as on datasets *hypo* and *german*, then the permutation-based method is preferred.

- The holdout approach is more conservative and more costly than the direct adjustment approach. The direct adjustment approach has already been criticized for inflating the number of false negatives unnecessarily [15]. Hence we do not recommend the use of the holdout approach.

During our experiments, we found that the interaction among frequent patterns is a big problem. If rule $R : X \to c$ is real and is statistically significant, then rules $X' \Rightarrow c$ are likely to be significant too, where $X'$ is a sub-pattern or super-pattern of $X$. This makes it very hard to determine what is a false positive. We use a simple method to tackle this problem. More sophisticated methods are needed.

Frequent patterns have a lot of redundancy among them. If the support of two patterns, $X$ and $X'$, is very close and $X$ is a sub-pattern of $X'$, then the two rules, $X \Rightarrow c$ and $X' \Rightarrow c$, are essentially testing the same hypothesis. It is desirable to reduce the redundancy and retain a small number of representative patterns for testing. This way, the number of tests is reduced and the power of the correction approaches can be improved. This will be our future work.

## 8. ACKNOWLEDGMENT

This work is supported in part by Singapore Agency for Science, Technology and Research grant SERC 102 101 0030.

## 9. REFERENCES

[1] H. Abdi. Bonferroni and Šidák corrections for multiple comparisons. *Encyclopedia of Measurement and Statistics. Thousand Oaks, CA: Sage*, 2007.

[2] R. Agrawal, T. Imielinski, and A. N. Swami. Mining association rules between sets of items in large databases. In *SIGMOD*, pages 207–216, 1993.

[3] S. D. Bay and M. J. Pazzani. Detecting group differences: Mining contrast sets. *Data Mining and Knowledge Discovery*, 5:213–246, 2001.

[4] Y. Benjamini and Y. Hochberg. Controlling the false discovery rate: a practical and powerful approach to multiple testing. *Journal of the Royal Statistical Society*, 57(1):125–133, 1995.

[5] S. Brin, R. Motwani, and C. Silverstein. Beyond market baskets: Generalizing association rules to correlations. In *SIGMOD Conference*, pages 265–276, 1997.

[6] G. E. Dallal. Historical background to the origins of p-values and the choice of 0.05 as the cut-off for significance. *(http://www.jerrydallal.com/LHSP/p05.htm)*, 2007.

[7] S. Dudoit and M. J. van der Laan. *Multiple Testing Procedures with Applications to Genomics*. Springer, 2008.

[8] R. A. Fisher. On the interpretation of $\chi^2$ from contingency tables, and the calculation of p. *Journal of the Royal Statistical Society*, 85(1):87–94, 1922.

[9] L. Geng and H. J. Hamilton. Interestingness measures for data mining: a survey. *ACM Computing Surveys*, 38(3), 2006.

[10] A. Kirsch, M. Mitzenmacher, A. Pietracaprina, G. Pucci, E. Upfal, and F. Vandin. An efficient rigorous approach for identifying statistically significant frequent itemsets. In *PODS*, pages 117–126, 2009.

[11] B. Liu, W. Hsu, and Y. Ma. Integrating classification and association rule mining. In *SIGKDD*, pages 80–86, 1998.

[12] G. Liu, H. Lu, and J. X. Yu. Cfp-tree: A compact disk-based structure for storing and querying frequent itemsets. *Information Systems*, 32(2):295–319, 2007.

[13] N. Megiddo and R. Srikant. Discovering predictive association rules. In *SIGKDD*, pages 274–278, 1998.

[14] N. Pasquier, Y. Bastide, R. Taouil, and L. Lakhal. Discovering frequent closed itemsets for association rules. In *ICDT*, pages 398–416, 1999.

[15] T. V. Perneger. What's wrong with bonferroni adjustments. *British Medical Journal*, 316:1236–1238, 1998.

[16] R. Rymon. Search through systematic set enumeration. In *Proceedings of the Internation Conference on Principles of Knowledge Representation and Reasoning*, pages 268–275, 1992.

[17] P.-N. Tan, V. Kumar, and J. Srivastava. Selecting the right interestingness measure for association patterns. In *SIGKDD*, pages 32–41, 2002.

[18] G. I. Webb. Discovering significant patterns. *Machine Learning*, 68(1):1–33, 2007.

[19] G. I. Webb. Layered critical values: A powerful direct-adjustment approach to discovering significant patterns. *Machine Learning*, 71(2-3):307–323, 2008.

[20] P. H. Westfall and S. S. Young. *Resampling-based multiple testing: examples and methods for P-value adjustment*. Wiley, 1993.

[21] X. Yin and J. Han. Cpar: Classification based on predictive association rules. In *SDM*, 2003.

[22] M. J. Zaki and K. Gouda. Fast vertical mining using Diffsets. In *SIGKDD*, 2003.